\documentstyle[aps,epsf,twocolumn]{revtex}
\begin{document}
\draft

\title{ Analog Electromagnetism in a Symmetrized $^3$He-A }

\author{ Jacek Dziarmaga }

\address{ Los Alamos National Laboratory,
          Theory Division T-6, 
          Los Alamos, New Mexico 87545, USA \\
          and
          Intytut Fizyki Uniwersytetu Jagiello\'nskiego,
          Reymonta 4, 30-059 Krak\'ow, Poland }
\date{ December 17, 2001 }
\maketitle

\begin{abstract}
We derive a low temperature effective action for the order parameter in a
symmetrized phase A of helium 3, where the Fermi velocity equals the
transversal velocity of low energy fermionic quasiparticles. The effective
action has a form of the electromagnetic action. This analog electromagnetism
is a part of the program to derive analog gravity and the standard
model as a low energy effective theory in a condensed matter system.
For the analog gauge field to satisfy the Maxwell equations interactions
in $^3$He require special tuning that leads to the symmetric case.  
\end{abstract}

\pacs{PACS numbers: 11.15.-q, 67.57.-z, 67.57.Jj, 11.90.+t}


\section{ Introduction }

  Quantum mechanics is not compatible with general relativity. There are well
known problems with the definition of a time operator, the cosmological constant
problem, the divergencies in the relativistic quantum field theory and the black
hole paradoxes \cite{hawking}. In recent years, as a result of interaction
between the condensed matter and the high energy physics communities, a new
program is emerging that may solve all these fundamental problems at once
\cite{review,casimir,bh,sakharov}. The key idea is that both general relativity
and the Standard Model \cite{review,casimir,bh,sakharov} are low energy effective
field theories of an underlying condensed matter system.

  By its very definition this program invalidates the time operator
problem. The fundamental theory is a condensed matter system described by
a ``nonrelativistic" $N$-body Schr\"odinger equation in an abstract
configuration space. The time operator problem is an artifact of the
effective low energy relativistic theory. When the missing definition of
a time operator leads to paradoxes in the effective theory, their
resolution can be found at the level of the fundamental condensed matter
system. 

  The cosmological constant or Casimir energy, when calculated within the
effective relativistic quantum field theory, is divergent. It is
customary to cut off this divergence at the Planck scale. Even when cut
off the cosmological constant is still, by many orders of magnitude,
inconsistent with observations. A fundamental condensed matter theory
should, at the very least, provide a correct prescription how to make the
cut off in the effective relativistic theory \cite{casimir}. An example
in Ref.\cite{casimir} demostrates that in a condensed matter system it is
even possible to have a nonzero Casimir force but at the same time an
exactly vanishing cosmological constant.

  The divergencies in the perturbative relativistic quantum field theory
are yet another artifact of the effective low energy theory. The
underlying $N$-body Schr\"odinger equation does not suffer from any
divergencies.

  Violation of relativity at high energies or strong fields, where the
``nonrelativistic" nature of the fundamental theory shows up, allows the high
energy particles to communicate over a black hole event horizon and in this way
it solves the paradoxes related to the horizon \cite{bh}.

  The idea that relativistic fields are low energy excitations of a
condensed matter system is older than the relativistic fields themselves.
Maxwell derived his famous equations as a hydrodynamic description of a
hypothetical ether. Later on the Michelson-Morley experiment proved that
there is no detectable motion of the Earth with respect to the ether. The
fundamental condensed matter system is an ether but in a modern guise.
There is an essential difference with respect to the traditional ether:
now ``everything", i.e. both light and fermionic matter (including the
famous Michelson and Morley's experimental setup), are effective low
energy bosonic and fermionic relativistic excitations. The low energy
relativistic excitations cannot detect their motion with respect to the
modern ether and there are no fundamental relativistic fields, otherwise
we would have to deal again with the time operator and the cosmological
constant problem. The condition that the low energy excitations must
include relativistic fermionic quasiparticles strongly suggests that the
underlying condensed matter system must contain fundamental
``nonrelativistic" fermions.

  Analogies between the black hole horizon and sonic horizons in a number of
condensed matter systems were explored in Refs.\cite{analogies}. Analogies
between {\it fermionic} helium 3 \cite{VW,GEV} and the standard model plus
general relativity were explored in depth by Volovik in Ref.\cite{review}. Of
particular interest in the present context are two phases of the superfluid
helium 3: the A phase and the planar phase \cite{review}. In a conventional
superconductor and in the $B$ phase of helium 3 there is an energy gap $\Delta_0$
between the Landau quasiparticles below the Fermi surface, where $p=p_F$, and
those above the Fermi surface. In the A phase and the planar phase this gap has
two nodes at the so called Fermi points on the Fermi surface. Order parameter
includes a unit vector $\hat{\bf l}$ related to orbital angular momentum of the
atoms. The two Fermi points are located at ${\bf p}=\pm p_F\hat{\bf l}$. Close to
the Fermi point, say, ${\bf p}=p_F\hat{\bf l}$ the energy $e_{\bf p}$ of the
fermionic Landau quasiparticles can be approximated by

\begin{equation}
e_{\bf p}^2\;+\;
g^{ab}(p_a-p_Fl_a)(p_b-p_Fl_b)\;
\approx\; 0,
\label{e2}
\end{equation}
where the indices $a,b$ run over $1,2,3$ (or $x,y,z$). This spectrum is
relativistic, there are low energy effective Dirac fermions in this
system. 

  In general, the metric tensor depends on $\hat{\bf l}$,

\begin{eqnarray}
&&
g^{00}\;=\;1\;,\\
&&
-g^{ab}\;=\;
c_F^2\; l^al^b\;+\;
c_{\bot}^2\; (\delta^{ab}-l^al^b)\;,
\end{eqnarray}
where $c_F$ is an effective Fermi velocity and $c_{\bot}=\Delta_0/p_F$ is
a transversal velocity of the fermionic quasiparticles near a Fermi
point. However, in a {\it symmetric} case, when

\begin{equation}
c_{\bot} \;=\; c_F \;, 
\end{equation}
the metric tensor becomes independent of $\hat{\bf l}$,

\begin{equation}
g^{\mu\nu}=
{\rm diag}\left\{1,-c_F^2,-c_F^2,-c_F^2\right\}\;,
\end{equation}
and $c_F$ becomes an effective velocity of light for the Dirac fermions.

  As noted in Ref.\cite{review} the $p_F\hat{\bf l}$ in Eq.(\ref{e2}) can
be interpreted as an electromagnetic vector potential and integration
over the relativistic fermions should give an effective electromagnetic
action for these gauge field. This integration over an equilibrium low
temperature ensamble of fermions is a subject of the next Section. This
derivation shows how an effective relativistic electrodynamics emerges
from an underlying fermionic condensed matter system.

  The derivation in the next Section generalizes the classic helium 3 results for
$c_F\gg c_{\bot}$ obtained by Cross in Ref.\cite{cross}. The symmetric case
$c_F=c_{\bot}$ is far from the real helium 3. However, it should be possible to
construct an abstract symmetrized helium 3 with interactions tuned so as to have
a stable phase A and $c_F=c_{\bot}$ at the same time. The fundamental condensed
matter system does not need to be constrained by the generic properties of
interactions in the electronic or atomic condensed matter systems. The aim of
this paper is to better substantiate the idea \cite{review} that the {\it
relativistic} electrodynamics can be an effective low energy theory in a ``{\it
nonrelativistic}" fermionic condensed matter system.


\section{ The effective electromagnetism }


\subsection{ Bogolubov-Nambu space}

  To describe helium 3 it is convenient to combine spin-up and spin-down
fermions into a Bogolubov-Nambu spinor

\begin{equation}
\chi({\bf x})=
\pmatrix{ \psi_{\uparrow}({\bf x})              \cr 
          \psi_{\downarrow}({\bf x})            \cr 
          \psi_{\downarrow}^{\dagger}({\bf x})  \cr 
          -\psi_{\uparrow}^{\dagger}({\bf x})   \cr}\;.
\label{chi}
\end{equation}
It is understood here that ${\bf p}=-i\nabla$ and the nabla is applied to
the $\chi({\bf x})$ on the right. A mean field Hamiltonian that describes
interaction of the fermionic atoms with the order parameter in the phase
A of $^3$He is given by

\begin{equation}
H=\frac12\int d^3x\;
\chi^{\dagger}({\bf x})
\pmatrix{  +\epsilon_{\bf p}                       & 
           \Delta_0^*\sigma\frac{p_{\bot}^*}{p_F}  \cr
           \Delta_0\sigma\frac{p_{\bot}}{p_F}      &
           -\epsilon_{\bf p}                       \cr }
\chi({\bf x})
\label{H}
\end{equation}
Here $\Delta_0({\bf x})$ is the energy gap and $p_F$ is the Fermi
momentum. $\epsilon({\bf p})$ is a quasiparticle energy, which can be
approximated close to the Fermi surface by

\begin{equation}
\epsilon_{\bf p}\;\approx\;
\frac{p^2}{2m_*}-\frac{p^2_F}{2m_*}\;=\;
\frac{(p+p_F)(p-p_F)}{2m_*}
\;\approx\;
c_F\;(p-p_F)\;,
\label{epsilon}
\end{equation}
where $c_F=p_F/m_*$ is a Fermi velocity and $m_*$ is an effective mass
of Landau quasiparticles close to the Fermi surface.

\begin{equation}
\sigma({\bf x})\equiv d^{\mu}({\bf x})\sigma_{\mu}
\end{equation}
with $d^{\mu}d^{\mu}=1$ is a $2\times 2$ spin matrix. 

\begin{equation}
p_{\bot}({\bf x}) \equiv
\frac12
\{
e^a_1({\bf x})+ie^a_2({\bf x}),
p^a
\}\;,
\end{equation}
where summation runs over $a=1,2,3$, and  $\hat{\bf e}_1$ and
$\hat{\bf e}_2$ satisfy

\begin{equation}
\hat{\bf e}_1\hat{\bf e}_1=1 \;,
\hat{\bf e}_2\hat{\bf e}_2=1 \;,
\hat{\bf e}_1\hat{\bf e}_2=0 \;,
\hat{\bf l}=\hat{\bf e}_1\times\hat{\bf e}_2 \;.
\end{equation}


\subsection{ Background order parameter }

 We will derive an effective action for small fluctuations of the order parameter
around the equilibrium order parameter

\begin{eqnarray}
&&
\Delta_0({\bf x})=\Delta_0\in{\cal R}\;, 
\label{b1}\\
&&
\sigma({\bf x})=\sigma_3\;,
\label{b2}\\
&&
\hat{\bf e}_1({\bf x})=\hat{\bf e}_x\;
\label{b3}\\
&&
\hat{\bf e}_2({\bf x})=\hat{\bf e}_y\;
\label{b4}\\
&&
\hat{\bf l}({\bf x})=\hat{\bf e}_x\times\hat{\bf e}_y=
\hat{\bf e}_z\;
\label{b5}\\
&&
p_{\bot}=p_x+ip_y\;.
\label{b6}
\end{eqnarray}
With this background the Hamiltonian (\ref{H}) becomes

\begin{equation}
H_0=\frac12\int d^3x\;
\chi^{\dagger}({\bf x})
\pmatrix{  +\epsilon_{\bf p}                       &
           \Delta_0\sigma_3\frac{p_{\bot}^*}{p_F}  \cr
           \Delta_0\sigma_3\frac{p_{\bot}}{p_F}    &
           -\epsilon_{\bf p}                       \cr }
\chi({\bf x})\;. 
\label{H0}
\end{equation}


\subsection{ Bogolubov transformation }

The Hamiltonian (\ref{H0}) is diagonalized by a Bogolubov transformation

\begin{eqnarray}
&&
\psi_{\uparrow}({\bf p})=
u_{\bf p}
\gamma_{\uparrow}({\bf p})+
v_{\bf p}
\gamma_{\downarrow}^{\dagger}(-{\bf p}) \;,\\
&& 
\psi_{\downarrow}({\bf p})=
u_{\bf p}
\gamma_{\downarrow}({\bf p})+
v_{\bf p}
\gamma_{\uparrow}^{\dagger}(-{\bf p}) \;,
\label{gamma}
\end{eqnarray}
where the Bogolubov coefficients $u_{\bf p}$ and $v_{\bf p}$ satisfy

\begin{eqnarray}
&&
|u_{\bf p}|^2=
\frac12
\left(
1+\frac{\epsilon_{\bf p}}{e_{\bf p}}
\right) \;,\\
&&
|v_{\bf p}|^2=
\frac12
\left(
1-\frac{\epsilon_{\bf p}}{e_{\bf p}}
\right)\;,\\
&&
2u_{\bf p}v_{\bf p}=
\frac{c_{\bot}p_{\bot}}{e_{\bf p}} \;,\\
&&
e_{\bf p}=
\left(
\epsilon_{\bf p}^2+
c_{\bot}^2
|p_{\bot}|^2
\right)^{1/2}
\label{e}
\end{eqnarray}
Here we define $c_{\bot}\equiv \Delta_0/p_F$. The diagonalized Hamiltonian
(\ref{H0}) is

\begin{equation}
H_0=\int d^3p\;
e_{\bf p}
\left[
\gamma_{\uparrow}^{\dagger}({\bf p})
\gamma_{\uparrow}({\bf p})+
\gamma_{\downarrow}^{\dagger}({\bf p})
\gamma_{\downarrow}({\bf p})
\right]\;\;.
\end{equation}
Close to the Fermi point at ${\bf p}=\pm p_F\hat{\bf l}$ the energy
squared of the quasiparticles can be approximated by

\begin{equation}
e_{\bf p}^2\;+\;
g^{ab}(p_a-p_Fl_a)(p_b-p_Fl_b)\;\approx\;0,
\end{equation}
compare with Eqs.(\ref{epsilon},\ref{e}). $g^{ab}$ is a spatial part of a
metric tensor

\begin{equation}
g^{\mu\nu}=
{\rm diag}\left\{1,-c_{\bot}^2,-c_{\bot}^2,-c_F^2\right\}\;.
\label{g}
\end{equation}


\subsection{ Small fluctuations of $\hat{\bf l}$. }

  We add small perturbations to the background field
(\ref{b1},..,\ref{b4})

\begin{eqnarray}
&&
\hat{\bf e}_1({\bf x})=
\hat{\bf e}_x+{\bf n}_1({\bf x})\;,\\
&&
\hat{\bf e}_2({\bf x})=
\hat{\bf e}_y+{\bf n}_2({\bf x})\;
\end{eqnarray}
and define a small complex vector field

\begin{equation}
z_a({\bf x})\equiv
n^a_1({\bf x})+
i\; n^a_2({\bf x})\;.
\label{z}
\end{equation}
The Hamiltonian (\ref{H}) becomes $H=H_0+H_1+{\cal O}(z^2)$ with

\begin{equation}
H_1=
\int d^3p\;
\left[
z_a^*({\bf p})
F^a_{\bf p}+
{\rm h.c.}
\right] \;,
\label{H1}
\end{equation}
an interaction Hamiltonian linear in $z_a$. Here we use the Fourier
transform

\begin{equation}
z_a({\bf p})=
\int\frac{d^3x}{(2\pi)^3}\;
e^{-i{\bf x}{\bf p}}\;
z_a({\bf x}) \;,
\end{equation}
and the operator

\begin{eqnarray}
F^a_{\bf p}[&\gamma &]
\equiv
-\Delta_0
\int d^3k\;\; 
\frac{k^a}{p_F} \;\times
\nonumber\\
&\left[\right.&
u_{\frac{\bf p}{2}+{\bf k}}
u_{\frac{\bf p}{2}-{\bf k}}
\gamma_{\downarrow}(\frac{\bf p}{2}+{\bf k})
\gamma_{\uparrow}(\frac{\bf p}{2}-{\bf k})
+\nonumber\\
&&
v_{\frac{\bf p}{2}+{\bf k}}
v_{\frac{\bf p}{2}-{\bf k}}
\gamma_{\uparrow}^{\dagger}(-\frac{\bf p}{2}-{\bf k})
\gamma_{\downarrow}^{\dagger}(-\frac{\bf p}{2}+{\bf k})
+\nonumber\\
&&
u_{\frac{\bf p}{2}+{\bf k}}
v_{\frac{\bf p}{2}-{\bf k}}
\gamma_{\downarrow}(\frac{\bf p}{2}+{\bf k})
\gamma_{\downarrow}^{\dagger}(-\frac{\bf p}{2}+{\bf k})
+\nonumber\\
&&
\left.
u_{\frac{\bf p}{2}-{\bf k}}
v_{\frac{\bf p}{2}+{\bf k}}
\gamma_{\uparrow}^{\dagger}(-\frac{\bf p}{2}-{\bf k})
\gamma_{\uparrow}(\frac{\bf p}{2}-{\bf k})
\right] \;.
\label{F}
\end{eqnarray}


\subsection{ Second order effective action }

 A real (unitary) part of the second order effective action is

\begin{eqnarray}
&&
S^{(2)}[z]=\\
&&
{\rm Re}\;\frac{i}{2}\int dt\; dt'\;
\langle\hat{T}H_1[\gamma_+(t)]H_1[\gamma_+(t')] \rangle =
\nonumber\\
&&
{\rm Re}\;
\frac{i}{2}\int dt\; dt'
\int d^3p\; d^3p'\times
\label{S2z}
\nonumber\\
&&
\left[ 2 
z_a(t,{\bf p})\;
\right.
\langle\; 
\hat{T}\;
F^{\dagger a}_{\bf p}[\gamma_+(t)]\;
F^{b}_{\bf p'}[\gamma_+(t')]\;
\rangle\;
z_b^*(t',{\bf p'})+
\nonumber\\
&&
(z_a^*(t,{\bf p})\;
\left.
\langle\;
\hat{T}\;
F^{a}_{\bf p}[\gamma_+(t)]\;
F^{b}_{\bf p'}[\gamma_+(t')]\;
\rangle\;
z_b^*(t',{\bf p'})
+{\rm c.c.})
\right],
\nonumber
\end{eqnarray}
where $\hat{T}$ means time ordering along the Kyeldysh contour.  The interaction
picture $\gamma_+(t)$'s sit on the positive (forward in time) branch of the
contour. A straightforward but somewhat tedious calculation, which uses a
correlator time ordered along the contour

\begin{eqnarray}
\langle \hat{T}[
\gamma_+(t,{\bf k})
&& 
\gamma_+^{\dagger}(t',{\bf k}')
]\rangle\;=\;
\delta({\bf k}-{\bf k}')\;
e^{-ie_{\bf k}(t-t')} \times 
\nonumber\\
&&
[\theta(t-t')f(-\beta e_{\bf k})-
 \theta(t'-t)f(+\beta e_{\bf k}) ]
\end{eqnarray}
with $f(x)=(1+e^x)^{-1}$ and $\beta$ an inverse temperature, gives an
effective action

\begin{eqnarray}
S^{(2)}[z]=
\int d\omega
&&
\int d^3p\;\times
\\
\left[
z_a(\omega,{\bf p})\;
\right.
&&
G^{ab}_1(\omega,{\bf p})\;
z_b^*(\omega,{\bf p})+
\nonumber\\
(z_a^*(\omega,{\bf p})\;
&&
\left.
G^{ab}_2(\omega,{\bf p})\;
z_b^*(-\omega,-{\bf p})+
{\rm c.c.})
\right].
\nonumber
\end{eqnarray}
The kernels are given by 

\begin{eqnarray}
G^{ab}_1&&(\omega,{\bf p})\;=\;
2\pi\; \Delta_0^2 \;\;
{\rm P.V.}\;
\int d^3k\;
\frac{k^ak^b}{p_F^2}\times
\nonumber\\
&&
\frac{2\sinh(\beta e_{\bf k})}{1+\cosh(\beta e_{\bf k})}
\times
\nonumber\\
&&
\left[
\frac{|u_{{\bf k+\frac{p}{2} }}|^2\;|u_{{\bf k-\frac{p}{2} }}|^2}
     {+\omega+e_{{\bf k+\frac{p}{2} }}+e_{{\bf k-\frac{p}{2} }}}+
\frac{|v_{{\bf k+\frac{p}{2} }}|^2\;|v_{{\bf k-\frac{p}{2} }}|^2}
     {-\omega+e_{{\bf k+\frac{p}{2} }}+e_{{\bf k-\frac{p}{2} }}}
\right]
\end{eqnarray}
and

\begin{eqnarray}
G^{ab}_2&&(\omega,{\bf p})\;=\;
-\pi\; \Delta_0^2 \;\;
{\rm P.V.}\;
\int d^3k\;
\frac{k^ak^b}{p_F^2}\times
\nonumber\\ 
&&
\frac{2\sinh(\beta e_{\bf k})}{1+\cosh(\beta e_{\bf k})}
\times
(u_{\bf k+\frac{p}{2} }
 v_{\bf k+\frac{p}{2} })
(u_{\bf k-\frac{p}{2} }
 v_{\bf k-\frac{p}{2} })
\times
\nonumber\\
&&
\left[ 
\frac{1}{+\omega+e_{{\bf k+\frac{p}{2} }}+e_{{\bf k-\frac{p}{2} }}}+
\frac{1}{-\omega+e_{{\bf k+\frac{p}{2} }}+e_{{\bf k-\frac{p}{2} }}}
\right]
\end{eqnarray}
Here we neglect terms that are exponentially small for small temperature.
In order to get a low energy effective theory, these kernels will be
(gradient) expanded in powers of $\omega$ and ${\bf p}$.

\subsection{ Gradient expansion of $G^{33}$ }

  A gradient expansion of $G^{33}$ gives terms which are logarithmically
divergent when $\beta\to\infty$. This divergence, localized at the Fermi
points ${\bf k}=\pm p_F\hat{\bf l}$, can be identified as

\begin{eqnarray}
G&&^{33}_{1,\rm Log}(\omega,{\bf p})=
\nonumber\\
&&
\frac{4\pi^2\Delta_0^2}{3}
\left[
\omega^2-\frac12 c_{\bot}^2(p_x^2+p_y^2)-c_F^2(p_z^2)
\right]
\ln(\beta\Delta_0)\;.
\end{eqnarray}
and

\begin{equation}
G^{33}_{2,\rm Log}(\omega,{\bf p})=
\frac{\pi^2\Delta_0^2}{3}
\left[
c_{\bot}^2
p_{\bot}^2
\right]
\ln(\beta\Delta_0)\;.
\end{equation}
After inverse Fourier transform we obtain the logarithmically
divergent part of the second order effective action

\begin{eqnarray}
&&
S^{(2)}_{\rm Log}[{\bf n}]=
\frac{p_F^2  \ln\left( \frac{\Delta_0^2}{T^2} \right)   }{24\pi^2\;c_F}
\int d^4x\;
\times\label{Sn}\\
&&
\left\{
\sum_{k=1,2}
\left[
\left(
\frac{\partial n_k^3}{\partial t}
\right)^2-
c_F^2 
\left(
\frac{\partial n_k^3}{\partial z}
\right)^2
\right]^2-
c_{\bot}^2
\left[
\partial_x n_2^3 - \partial_y n_2^3
\right]^2
\right\}.
\nonumber
\end{eqnarray}
This action is a second order perturbative version of an action

\begin{eqnarray}
&&
S^{(2)}_{\rm Log}[{\bf l}]=
\frac{p_F^2  \ln\left( \frac{\Delta_0^2}{T^2} \right)  }{24\pi^2\;c_F}
\int d^4x\;
\times
\nonumber\\
&&
\left\{
\left[
\frac{\partial {\bf l}}{\partial t}
\right]^2-
c_F^2
\left[
{\bf l}\times({\bf\nabla}\times{\bf l})
\right]^2-
c_{\bot}^2
\left[
{\bf l}({\bf\nabla}\times{\bf l})
\right]^2
\right\}.
\label{Sl}
\end{eqnarray}
Fluctuations of $\hat{\bf l}$ are not the only contribution to the
logarithmically divergent part of the low energy effective action.
Another contribution comes from the component of the superfluid velocity
${\bf v}$ which is parallel to $\hat{\bf l}$.


\subsection{ Small fluctuations of $\left(\hat{\bf l}{\bf v}\right)$ }

  For a {\it uniform} stationary superfluid flow with velocity ${\bf v}$
and close to the Fermi surface, $p\approx p_F$, the Hamiltonian
(\ref{H0}) becomes

\begin{eqnarray}
H_0&=&\frac12\int d^3x\;
\chi^{\dagger}({\bf x}) \times\label{Hv}\\
&&
\pmatrix{  +\epsilon_{{\bf p}+m_*{\bf v}}+\frac12 m_*v^2           &
           \Delta_0\sigma_3\frac{(p_{\bot}^*+m_* v_{\bot}^*)}{p_F} \cr
           \Delta_0\sigma_3\frac{(p_{\bot}-m_* v_{\bot})}{p_F}     &
           -\epsilon_{{\bf p}-m_*{\bf v}}-\frac12 m_*v^2           \cr}
\chi({\bf x})\;,
\nonumber   
\end{eqnarray}
compare with Eqs.(\ref{H},\ref{epsilon}) and use a Galilean
transformation. Here $v_{\bot}\equiv v_x+iv_y$. We are interested in the
part of the Hamiltonian (\ref{Hv}) that is linear in ${\bf v}$ and we
expand
  
\begin{equation}
\epsilon_{{\bf p}+m_*{\bf v}}\;=\;
\epsilon_{\bf p}\;+\;
{\bf p}{\bf v}\;+\;
{\cal O}(v^2)\;.
\label{epsilonv}
\end{equation}
So far ${\bf v}$ was constant. Now we make it space and time dependent,
${\bf v}={\bf v}(t,{\bf x})$, and at the same time, to keep the
Hamiltonian (\ref{Hv}) hermitian, we make in Eq.(\ref{epsilonv}) a
replacement
  
\begin{equation}
{\bf p}{\bf v}\;\to\;
\frac12\left\{ {\bf p} , {\bf v}(t,{\bf x})  \right\}\;=\;
\frac12[{\bf p}{\bf v}(t,{\bf x})]+
{\bf v}(t,{\bf x}){\bf p}\;.
\label{replace}
\end{equation}
We expand the Hamiltonian (\ref{Hv}) to leading order in ${\bf v}$ using
Eq.(\ref{epsilonv}) and the replacement (\ref{replace}). In the expanded
Hamiltonian we keep only terms where the operator ${\bf p}$ is applied to
$\chi$ or $\chi^{\dagger}$. As the main contribution to the
logarithmically divergent part of the effective action comes from near
the Fermi points at ${\bf p}=\pm p_F\hat{\bf l}$, these terms are
formally of the order of $p_F$. They are large as compared to terms where
the operator ${\bf p}$ is applied to the slowly varying velocity field
${\bf v}$. After those last terms are neglected the interaction
Hamiltonian becomes

\begin{equation}
H_1\;\approx\;
\frac12 \int d^3x\;
\chi^{\dagger}({\bf x})\;
\left[ v^a(\bf x)p_a \right]\;
\chi({\bf x})\;.
\label{H1v}
\end{equation}  
This Hamiltonian is hermitian when we take into account that $p_a{\bf v}(\bf x)$
is negligible as compared to $p_a\chi^{(\dagger)}$. With the definition
(\ref{chi}) and the Bogolubov transformation (\ref{gamma}) the Hamitonian becomes

\begin{equation}
H_1\;\approx\;  
\int d^3p\;
\left[ v_a^*({\bf p})\;f^a_{\bf p} + {\rm h.c.} \right]   
\label{H1vv}
\end{equation}
where

\begin{eqnarray}
f^a_{\bf p}\;\equiv\;&&
\int d^3k\; k_a
\times\nonumber\\
&&
\left(
u_{{\bf k}+\frac{\bf p}{2}}
v^*_{{\bf k}-\frac{\bf p}{2}}-
u_{{\bf k}-\frac{\bf p}{2}}
v^*_{{\bf k}+\frac{\bf p}{2}}
\right)
\gamma_{\downarrow}(-{\bf k})\gamma_{\uparrow}(+{\bf k})
\label{f}\;.
\end{eqnarray}
Here we neglect all mixed terms of the form $\gamma^{\dagger}\gamma$
which for small $T$ give an exponentially small contribution to the
effective action. The effective action is given by

\begin{eqnarray}
S^{(2)}_{\rm Log}&&[{\bf v}]\approx
{\rm Re}\;
\frac{i}{2}\int dtdt'\int d^3pd^3p'\times
\nonumber\\
&&
2v_a(t,{\bf p})\;
\langle\;  
\hat{T}\;
f^{\dagger a}_{\bf p}[\gamma_+(t)]\;
f^{b}_{\bf p'}[\gamma_+(t')]\;
\rangle\;
v_b^*(t',{\bf p'})\;,
\label{S2v}
\end{eqnarray}
compare to Eq.(\ref{S2z}). A straightforward calculation similar to the
derivation of the effective action for small fluctuations of $\hat{\bf
l}$ gives

\begin{equation}
S^{(2)}_{\rm Log}({\bf v})=
\frac{p_F^2\ln\left(\frac{\Delta_0^2}{T^2}\right)}{24\pi^2}
\int\sqrt{-g}\; d^4x\;
\left(-g^{ab}\partial_a v_3 \partial_b v_3\right) \;.
\label{S2A0}
\end{equation}
The logarithmically divergent part of the effective action (\ref{S2A0}) contains
only $v_3$ because at the Fermi points it is only $v_3$ that couples to ${\bf
p}=\pm p_F\hat{\bf l}=\pm p_F\hat{\bf e}_z$, compare Eq.(\ref{H1v}).


\subsection{ The electromagnetic effective action }

  In the symmetric $^3$He-A,

\begin{eqnarray}
&& c_F=c_{\bot}\;,\nonumber\\
&& g^{\mu\nu}=
   {\rm diag}\left\{ 1,-c_F^2,-c_F^2,-c_F^2 \right\}\;,
\label{gsymm}
\end{eqnarray}
and after identifications

\begin{eqnarray}
&&
A_0     \;=\; p_F\left({\bf l}{\bf v}\right) \;,\nonumber\\
&&
{\bf A} \;=\; p_F\hat{\bf l}
\label{AA}
\end{eqnarray}
the sum of the two actions (\ref{Sl},\ref{S2A0}) becomes 

\begin{equation}
S^{(2)}_{\rm Log}\;=\;
\frac{\ln\left( \frac{\Delta_0^2}{T^2} \right)}{12\pi^2}\;
\int\sqrt{-g}\;d^4x\;
\left(
 -\frac14 F^{\mu\nu}F_{\mu\nu}
\right)\;,
\label{F2}
\end{equation}
where

\begin{equation}
F_{\mu\nu}=\partial_{\mu}A_{\nu}-\partial_{\nu}A_{\mu}\;.
\end{equation}
In the symmetric case the metric tensor (\ref{gsymm})does not depend any more on
the direction of $\hat{\bf l}$. $\hat{\bf l}$ can be unambiguously interpreted as
a vector potential. The symmetrization is essential for the interpretation of
$\hat{\bf l}$ as a vector potential.


\section{ Conclusion }

  The effective action (\ref{F2}) and the identifications \label{AA} agree with
the effective action and the identifications that were suggested in
Ref.\cite{review}.

  The relativistically invariant low temperature effective action comes from
integration near the Fermi points where the quasiparticles are well approximated
by relativistic Dirac fermions. The effective electromagnetic action plus Dirac
fermions minimally coupled to the gauge field give rise to an effective
relativistic electrodynamics emerging from an underlying nonrelativistic
fermionic condensed matter system.

  The dispersion relation in Eq.(\ref{e2}) allows one to interpret the
$p_F\hat{\bf l}$ as a gauge field for the relativistic fermions near a Fermi
point. However, for this gauge field to satisfy the gauge invariant Maxwell
equations of motion we need a right tuning of interactions in the model so that
$c_F=c_{\bot}$. Symmetry considerations based on Eq.(\ref{e2}) alone are not
enough to get the right dynamics for the gauge field.


\section*{ Acknowledgements }

I would like to thank Grisha Volovik for his contribution to this work, Pawel
Mazur for his informal one audience lectures, and Diego Dalvit for his comments
on an earlier draft of this manuscript.


\end{document}